\documentclass[%
 aip,
 amsmath,amssymb,
preprint,%
 reprint,%
]{revtex4-1}

\usepackage[]{graphicx,xcolor}
\usepackage{dcolumn}
\usepackage{bm}
\usepackage{textcomp}

\newcommand{\blue}{\textcolor{black}}
\usepackage[utf8]{inputenc}
\usepackage[T1]{fontenc}
\usepackage{mathptmx}

\newcommand{\aap}{{\it Astronomy \& Astrophysics }}
\newcommand{\procspie}{{\it SPIE}}

\begin{document}
\preprint{AIP/123-QED}
\title[]{\blue{Demonstration of MHz frequency domain multiplexing readout of 37 transition edge sensors for high-resolution X-ray imaging spectrometers}}
\author{H. Akamatsu}  \email{h.akamatsu@sron.nl}
 \altaffiliation[corresponding author~]{SRON Netherlands Institute for Space Research, Niels Bohrweg 4, 2333 CA Leiden, The Netherlands}
\author{D. Vaccaro}  \altaffiliation{SRON Netherlands Institute for Space Research, Niels Bohrweg 4, 2333 CA Leiden, The Netherlands}
\author{L. Gottardi}  \altaffiliation{SRON Netherlands Institute for Space Research, Niels Bohrweg 4, 2333 CA Leiden, The Netherlands}
\author{J. van der Kuur}  \altaffiliation{SRON Netherlands Institute for Space Research, 
Landleven 12, 9747 AD Groningen, The Netherlands}
\author{C.P. de Vries}   \altaffiliation{SRON Netherlands Institute for Space Research, Niels Bohrweg 4, 2333 CA Leiden, The Netherlands}
\author{M. Kiviranta} \altaffiliation{VTT, Tietotie 3, 02150 Espoo, Finland}
\author{K. Ravensberg}  \altaffiliation{SRON Netherlands Institute for Space Research, Niels Bohrweg 4, 2333 CA Leiden, The Netherlands}
\author{M. D'Andrea}\altaffiliation{
SRON Netherlands Institute for Space Research, Niels Bohrweg 4, 2333 CA Leiden, The Netherlands, and also 
Institute for Space Astrophysics and Planetology (INAF/IAPS Roma): Roma, Lazio, Italy}
\author{E. Taralli}  \altaffiliation{SRON Netherlands Institute for Space Research, Niels Bohrweg 4, 2333 CA Leiden, The Netherlands}
\author{M. de Wit}   \altaffiliation{SRON Netherlands Institute for Space Research, Niels Bohrweg 4, 2333 CA Leiden, The Netherlands}
\author{M.P. Bruijn}  \altaffiliation{SRON Netherlands Institute for Space Research, Niels Bohrweg 4, 2333 CA Leiden, The Netherlands}
\author{P. van der Hulst}   \altaffiliation{SRON Netherlands Institute for Space Research, Niels Bohrweg 4, 2333 CA Leiden, The Netherlands}
\author{R. H. den Hartog}  \altaffiliation{SRON Netherlands Institute for Space Research, Niels Bohrweg 4, 2333 CA Leiden, The Netherlands}
\author{B-J. van Leeuwen}   \altaffiliation{SRON Netherlands Institute for Space Research, Niels Bohrweg 4, 2333 CA Leiden, The Netherlands}
\author{A.J. van der Linden}  \altaffiliation{SRON Netherlands Institute for Space Research, Niels Bohrweg 4, 2333 CA Leiden, The Netherlands}
\author{A.J McCalden}  \altaffiliation{SRON Netherlands Institute for Space Research, Niels Bohrweg 4, 2333 CA Leiden, The Netherlands}
\author{K. Nagayoshi}  \altaffiliation{SRON Netherlands Institute for Space Research, Niels Bohrweg 4, 2333 CA Leiden, The Netherlands}
\author{A.C.T. Nieuwenhuizen}  \altaffiliation{SRON Netherlands Institute for Space Research, Niels Bohrweg 4, 2333 CA Leiden, The Netherlands}
\author{M.L. Ridder}   \altaffiliation{SRON Netherlands Institute for Space Research, Niels Bohrweg 4, 2333 CA Leiden, The Netherlands}
\author{S. Visser}   \altaffiliation{SRON Netherlands Institute for Space Research, Niels Bohrweg 4, 2333 CA Leiden, The Netherlands}
\author{P. van Winden}  \altaffiliation{SRON Netherlands Institute for Space Research, Niels Bohrweg 4, 2333 CA Leiden, The Netherlands}
\author{J.R. Gao}
\altaffiliation{SRON Netherlands Institute for Space Research, Niels Bohrweg 4, 2333 CA Leiden, The Netherlands}
\altaffiliation[and also ]{Optics Group, Department of Imaging Physics, Delft University of Technology, Delft, 2628 CJ, The Netherlands}
\author{R.W.M. Hoogeveen}
\altaffiliation{SRON Netherlands Institute for Space Research, Niels Bohrweg 4, 2333 CA Leiden, The Netherlands}
\author{B.D. Jackson} \altaffiliation{SRON Netherlands Institute for Space Research, 
Landleven 12, 9747 AD Groningen, The Netherlands}
\author{J-W.A. den Herder}   \altaffiliation{SRON Netherlands Institute for Space Research, Niels Bohrweg 4, 2333 CA Leiden, The Netherlands}

\date{\today}

\begin{abstract}
We report on the development and demonstration of a MHz frequency domain multiplexing (FDM) technology to read out 
arrays of cryogenic transition edge sensor (TES) X-ray microcalorimeters. In our FDM scheme, TESs are AC biased at different resonant frequencies in the low MHz range through an array of high-$Q$ LC resonators. The current signals of all TESs are summed at superconducting quantum interference devices (SQUIDs). 
We have  demonstrated multiplexing for a readout of 31 pixels using room temperature electronics, high-$Q$ LC filters and TES arrays developed at SRON, and SQUID arrays from VTT. We  repeated this on a second setup with 37 pixels. The summed X-ray spectral resolutions $@$ 5.9 keV are $\Delta E_{\rm 31 pix ~MUX}=2.14\pm0.03$ eV and $\Delta E_{\rm 37 pix ~MUX}=2.23\pm0.03$ eV. The demonstrated results are comparable with other multiplexing approaches. There is potential to further improve the spectral resolution and to increase the number of multiplexed TESs, and to open up applications for TES X-ray microcalorimeters.
\end{abstract}

\maketitle

\begin{figure*}
  \centering
  \begin{tabular}{c}
    \includegraphics[width=1\linewidth, keepaspectratio]{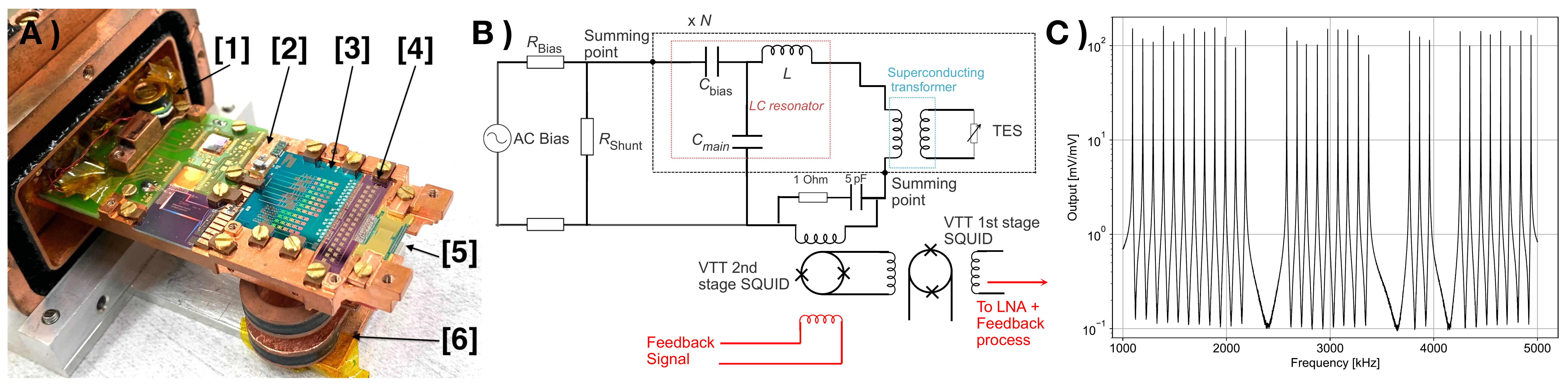} 
  \end{tabular}
  \caption{
  \label{fig:setup}
A) Photograph of the XFDM setup including [1] Ge thermistor, [2] VTT SQUIDs, [3] LC filter, [4] superconducting transformer, and [5] TES arrays. The Helmholtz coils [6] are placed between the TES arrays in the final assembly.
B) A schematic representation showing frequency domain multiplexing circuit.
$R_{\rm Bias}$, $R_{\rm Shunt}$, $C_{\rm Main}$ and $C_{\rm Bias}$ represent the load resistances, shunt resistances, capacitances for the LC resonator, and capacitances for the capacitive divider, respectively.
C)  32 LC resonators are measured by frequency-sweeping a small amplitude bias tone across the range of carrier frequencies from 1 to 5 MHz. The difference in the peak amplitude comes from the frequency resolution of the sweep. The scatter of the frequency separation is $\Delta f=98.8 \pm 2.0$ kHz. The scatter is due to intrinsic spread due to fabrication accuracy and parasitic inductance in the setup. 
All resonators have sufficient Q-factor to read out TES microcalorimeters.
  }
\end{figure*}

High-resolution X-ray spectroscopy is one of the most powerful techniques to understand the chemical composition of materials and the ionization state of plasma. It can be applied to a wide range of scientific fields varying from material science to the physics of hot plasma in the Universe\cite{2005cpd..book.....E}. 
Conventional X-ray spectrometers used in these applications, such as silicon based detectors or grating spectrometers, are sub-optimal: silicon based detectors have a high photon collecting area but moderate spectral resolution, while grating spectrometers have a high spectral resolution but a low collecting area and poor imaging capability. Cryogenic X-ray spectrometers have the potential of combining a high spectral resolution, imaging capability and photon collecting area by using a large number of pixel arrays.
X-ray microcalorimeters based on transition edge sensors (TESs)  are one of the most promising cryogenic spectrometer techniques for practical applications.  Recent progress of fabrication technologies makes the production of large format TES arrays possible. This has created the incentive to develop methods to read out up to several hundred TESs in one amplifier chain with low power dissipation.

One feasible signal multiplexing technology is frequency domain multiplexing (FDM)\cite{kiviranta02}.
Together with other technologies such as time division multiplexing\cite{durkin19}, code division multiplexing\cite{morgan16} and micro-wave multiplexing\cite{nakashima20}, the development of the FDM technology is progressing dramatically.
Under the FDM scheme, TES pixels are individually connected in series to high-$Q$ LC resonators with different frequencies and 
AC biased by a comb of frequencies in the MHz range. Signals from individual TESs are summed and read out by a SQUID  amplifier. The signals are amplified by a room temperature low noise amplifier (LNA) and subsequently digitized. 
For space flight applications,  the FDM readout has advantages\cite{vanderkuur16} of low power dissipation at cryogenic temperatures, individual adjustment of the TES operation points, ability to specify requirements of the superconducting quantum interference devices (SQUIDs), and light-weight shielding. We are developing FDM readout technology for astronomy in space,
such as the X-ray Integral Field Unit (X-IFU\cite{xifu18}) instrument on board the European X-ray astronomical satellite Athena,
and for future astronomical projects\cite{hubs, hazumi20, sdios}. For the Athena X-IFU instrument, 34 pixels within a frequency band of 1--5 MHz and spectral resolution $\Delta E = 2.5 $ eV $@$ 7 keV are required. In this letter, we report the recent progress for FDM readout of 37 pixels of TES X-ray microcalorimeters. 

For the FDM readout demonstration reported here, 
we used two setups, called \textit{XFDM} (Figure \ref{fig:setup} A) and \textit{40-pixel}. The names come simply from the development work at the time and we concentrate here on the XFDM setup.
The setups are mounted on the mixing chamber of a dry dilution cooler. Room temperature high-u metal and cryogenic superconducting Nb shields are employed to reduce the impact of environmental magnetic fields. The residual magnetic field, typically less than 0.2 uT,  is cancelled out by the superconducting Helmholtz coil. A germanium thermistor (Lake Shore GR50) is used to read out and control the setup temperature.  During the measurement, the setups were kept at 50 mK and 40 mK for XFDM and 40-pixel setups,  respectively. The typical temperature stability  is  < 1 \textmu K$_{\rm rms}$. 

Since both setups contain almost identical cryo components except for the number of resonators, we introduce here only 
the details of the XFDM setup. Figure~\ref{fig:setup} B shows the schematic diagram of the cryo-electronics.  The lithographically-made high-\textit{Q} LC filters contain parallel plate Nb/a-Si:H/Nb capacitors and Nb-based gradiometric spiral coils\cite{LCfilter18}. 
The inductor value is fixed, and the resonance frequencies are defined by the capacitor values.
In this report, we used 32 LC resonator chips with 2 uH  coils  (40 resonators for the 40-pixel setup) and resonances between 1 and 5 MHz with 100 kHz separation~(Fig.\ref{fig:setup} C). A shunt resistance $R_{\rm Shunt}$ (0.75 $\Omega$) is implemented at the 50 mK stage. As shown in Fig.\ref{fig:setup} middle, the LC filter contains a capacitive bias voltage divider with a ratio of 1:25, which makes an effective shunt resistance at the TES side to be 1.1 m$\Omega$. The measured \textit{Q}-factor when the TES is in the superconducting state is around $Q\sim 16\,000 \times(f/\rm [MHz])$,  limited by the effective shunt resistance in the circuit.  To be able to optimize the bandwidth of different TES designs, a superconducting transformer per pixel is employed. 
In this work, a transforming turns ratio of $n=1:1.125$  with a coupling factor   $k=0.94$ is used.
Under this configuration, TES are loaded by the effective inductance of 3.6 \textmu H.

\begin{figure}[t]
  \centering
        \includegraphics[width=1\linewidth, keepaspectratio]{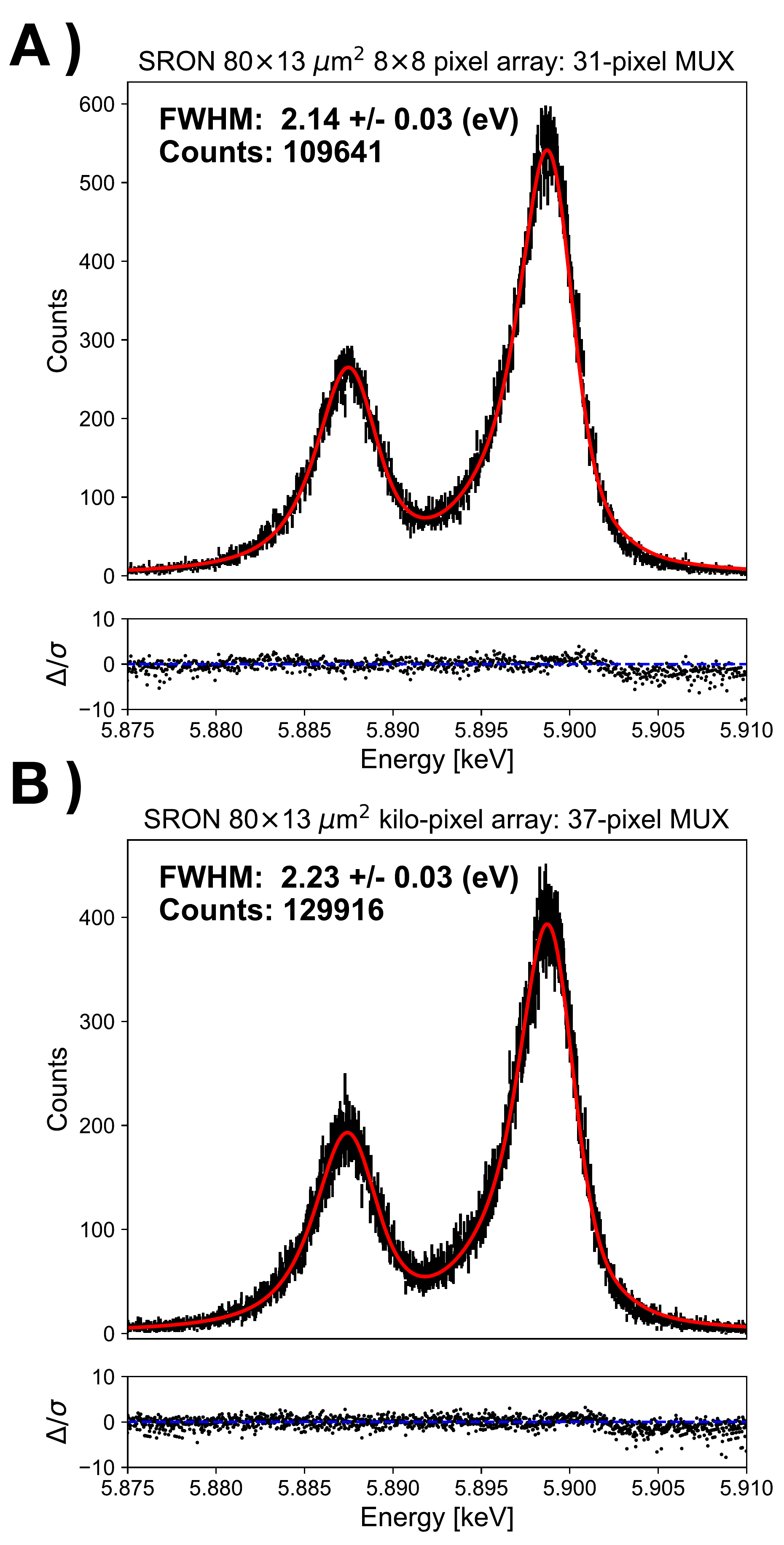} 
\caption{\label{fig:MUX_sum}
 The combined spectrum of 31 (A) and 37 (B) TESs simultaneous measurements using FDM readout.
The residual at the high energy flank is due to summing up of different resolution spectra.
}
\end{figure}

The SQUIDs-based amplifier is the most important component to read out a large number of TESs. 
We are using two-stage SQUID amplifiers, developed at VTT and which consist of 6 loops first stage and 184 $\times$ 4 loops 2nd stage SQUIDs\cite{kiviranta18, kiviranta21}.
Together with a SRON-developed low noise amplifier (LNA)\cite{wang20}, the typical readout current noise, referred to the SQUID input, is $\sim5~\rm pA/\sqrt{\rm Hz}$\cite{gottardi15}, which is much lower than the typical TES noise ($\sim$30-50 $\rm pA/\sqrt{\rm Hz}$). To avoid any undesired behaviour of the SQUIDs due to the coupling between the input coil of the SQUIDs, and parasitic inductance and capacitance in the LC filter\cite{vanderkuur16}, 
an RC filter with components $R=1~ \Omega$ and $C$ = 5 pF is implemented just before the input coil of  the first stage SQUID.

In our system, the multiplexed data is acquired with 20 M samples/s and the de-multiplexed data streams are decimated down to 156 k samples/s by using 4 stage filters on an SRON "DEMUX" digital board.  The DEMUX board,  using  AD9726 DACs and a Xilinx XC7V585T Virtex 7 FPGA,  
generates a modulated comb of AC bias frequencies for the TESs and the signals for the base-band feedback loop.
To increase the linearity and dynamic range of the SQUIDs,  the phase delay due to the harness between the room temperature electronics and cold electronics is compensated by the baseband feedback\cite{BBFB}.
The TES signals are demodulated and re-modulated with a controlled phase shift by the FPGA on the DEMUX board before being fed back to the system. The details of our digital system can be found in den Hartog et al. 2012 and 2018\cite{BBFB, denhartog18_DAC}.

For the FDM demonstration reported here, we employed detector arrays fabricated at SRON\cite{nagayoshi20}. 
They consist of a high-aspect ratio TES bilayer 80$\times$13 \textmu m of Ti (35 nm) and Au (200 nm), located on a 0.5 \textmu m thick Si$_{\rm x}$N$_{\rm y}$ membrane and coupled to a 240$\times$240~\textmu m Au absorber.
The elongated shape towards the direction of current flow was employed  to suppress the frequency dependent non-linearity\cite{gottardi18} by increasing TES resistance.
The thickness of the absorber is 2.35 \textmu m to ensure high quantum efficiency (83\% at 6 keV).
The critical temperature ($T_c$) of the bilayer is tuned to be around 83 mK and the thermal conductance at $T_{c}$ is $G\sim$65 pW/K.  The heat capacity is designed to be $C\sim$ 0.85 pJ/K. To reduce frequency dependent detector behaviour under AC bias, we use high aspect ratio TESs. 
This device was extensively investigated under AC bias with a lower inductance ($L_{\rm eff}$=1~\textmu H) and demonstrated an excellent spectral resolution of 1.8 eV FWHM\cite{dewit20}.

To characterise the performance of the TES, we used a $^{55}$Fe source located on the cryogenic platform.
The count rate was typically 1 count/s/pixel.  Each pulse was processed by the optimal filtering process\cite{optimalfiltering}  in the frequency space by using the time average pulse weighted by noise spectra.  Long-term gain drift correction was performed based on TES baseline current and pulse height information.  The energy non-linearity was corrected by using the zero energy (0 keV), Mn-K${\alpha}$ (5.9 keV) and Mn-K${\beta}$ (6.5 keV) information. The resultant energy spectrum was fitted with the  Mn-K${\alpha}$ line model\cite{holzer} convolved with the detector resolution. To avoid fitting bias, we employed Cash statistics~\cite{cash79, kaastra16}.

Previous demonstrations with FDM readout of TES spectrometers was mainly limited by three effects\cite{akamatsu20}. These are 1.) degradation of energy resolution due to carrier leakage from neighbouring resonators, 2.) the degradation due to inter-modulation distortion from the DACs, and 3.) frequency dependent non-linearity due to the Josephson junction in TESs under MHz AC bias. We solved or mitigated these effects using the following approaches. 

Carrier leakage from neighbouring pixels degrades the performance. The origin of this issue is that when the electrical bandwidth ($R/L$) of the devices is too large with respect to the frequency spacing, TESs become sensitive to the AC bias voltage of neighbouring frequencies. Therefore it is different from the electrical cross-talk due to parasitic and common inductance and capacitance. In the carrier leakage case, the TES voltage is modulated by $f_{\rm target}-f_{\rm nei}$, where $f_{\rm target}$ is a bias frequency of the target pixel and $f_{\rm nei}$ are the neighbouring frequencies. Although it could be avoided by applying large frequency separation or applying a phase window\cite{vanderkuur03, vanderkuur04}, these approaches would not be a practical solution in a real instrument. Instead, we solved the issue of excessive electrical bandwidth by increasing the inductance of the LC filter. However, too large an inductance will lead to an instability due to the electrical-thermal feedback\cite{irwin95} at the critical inductance. Moreover, large $L$ can lead to worse performance due to non-linear effects such as a large excursion in the resistance-temperature space (e.g., Fig. 4 of Kilbourne et al. 2008\cite{kilbourne08_MUX}).
The critical inductance can be estimated using the equation\,(7) in Smith et al. 2016\cite{smith16} showing strong dependency on the detector thermal time constant ($\sim C/G$), sensitivity to temperature ($\alpha$) and current ($\beta$) dependency.  
Previously these optimizations were hampered due to different $\alpha, \beta$ and the rather high $G$ of the available devices.
In this report, we used a factor 5 narrower electrical bandwidth configuration compared to the previous work\cite{akamatsu20}. 
The improvement is realized by the combination of a slower detector time constant and a different $\alpha-\beta$ relationship\cite{dewit20}.
Typical detector time constants are rise time $\tau_{\rm rise}\sim180$ \textmu s and fall time $\tau_{\rm fall}\sim1.1$ ms.

Inter-modulation products from non-linearity in the system are an issue for the detector performance and system dynamic range. In this report we refer to such products as inter-modulation line noise (IMLN). Since  more than 30 carriers are generated by the DAC, 
IMLN will be an additional noise term when it falls into the detector thermal response band ($<1 \rm kHz$).
This IMLN will degrade the performance and limit the number of multiplexed pixels.
This issue can be avoided by using carriers in a frequency arrangement where the frequency spacing between all subsequent carriers are on a regular grid\cite{denhartog18_DAC} (called grid-frequency). 
However, this will be a trade-off with the performance degradation due to an additional (virtual) shunt resistance resulting from operating off-resonance ($2\times2\pi L \Delta f$)\cite{akamatsu20}. Furthermore, LC filter fabrication accuracy puts a limit on the attainable resonance frequency distribution. To overcome the problem, we developed a frequency shift algorithm, which allows us to shift a resonator digitally without losing the performance.  The basic concept and practical demonstration are given in previous works\cite{vanderkuur18, akamatsu20}.
The multi-pixel behaviour was characterized carefully and we concluded that we could operate the system with more than 40 pixels, without the current DAC going into saturation\cite{vaccaro21_FSA, van_der_hulst_frequency_2021}.  In this report we used the grid frequency 1.0-2.0 kHz, which is well above the thermal bandwidth.

\begin{table}[th]
\caption{Summary of the FDM demonstrations}
\begin{center}
\begin{tabular}{lccccccccc} \hline \hline
Setup	&	N$_{\rm res}$ 	&  N$_{\rm MUX}$		& 	$\textless\Delta E_{\rm Single}> $  &	 $\Delta E_{\rm MUX}$	& $\Delta E_{\rm deg}$ \\  
		&				&					& 	[eV]							& 	[eV]				&	[eV]	\\
\hline
XFDM	&	32			&	31				&	1.95 						&	2.14$\pm0.03$ 	& 0.9		\\
40-pixel	&	38			&	37				&	2.05 						&	2.23$\pm0.03$ 	& 1.0		\\
\hline
\multicolumn{6}{l}{
$\Delta E_{\rm deg} = \Delta E_{\rm MUX}^2-\Delta E_{\rm single}^2$
}\\
\multicolumn{6}{l}{
In the XFDM setup, 1 resonator is affected by multiple IMLN.
} \\
\multicolumn{6}{l}{
In the 40-pixel setup, 1 resonator is affected by multiple IMLN } \\
\multicolumn{6}{l}{
and 2 resonators are missing due to fabrication yield.
}
\end{tabular}
\end{center}
\label{tab:mux}
\end{table}%
\blue{A frequency dependent non-linear behaviour under MHz AC bias} was observed in previous demonstrations\cite{akamatsu14_ACBias, gottardi14a, gottardi14}. 
The issue results from the so-called  proximity effect\cite{sadleir10, sadleir11}, 
where the local order parameter of the bilayer is modified by the connection to the niobium leads. The resulting structure shows similar behaviour to a weak superconducting link.
Under MHz AC bias, this effect shows several characteristics such as non-linear response, a Fraunhofer-like dependence of the critical current on magnetic field and steps in the superconducting transition. The resulting negative effect on the energy resolution scales with the resistance of the TES and is higher for low-ohmic TESs.
By comparing devices with different impedance and 
saturation power, 
the power needed to drive TES into the normal state,  we confirmed 
we confirmed our prediction\cite{gottardi18} that the Josephson current decreases with the increase of the superconducting phase difference over the TES, $\Phi\sim\frac{\sqrt{PR}}{\omega}$, where 
$P$ is the detector power, $R$ is the resistance, and $\omega$ is the bias frequency.
Accordingly,  high-power and high-resistance devices show the least degradation of the energy resolution.
Since the detector power is defined by the application, we modified the detector resistance by changing the aspect ratio of the TESs. 
Despite the large normal resistance $R_{\rm N}$,  we have shown that the internal thermal fluctuation noise
  remains small for these devices\cite{gottardi21}. Moreover, a high aspect ratio TES minimizes the impact of the loss due to eddy  currents. The average single pixel performance over 1--5 MHz $\textless\Delta E_{\rm Single}> $ is given in Table~\ref{tab:mux}, and is slightly worse than one with small inductance measurements ( $\Delta E_{\rm small~\it{L}}\sim$1.8 eV).  
This is likely due to the large inductance, and the non-linearity and large signal effects associated with that..
For the demonstration setups, a 8$\times$8 TES array is implemented for the XFDM and a 32$\times$32 TES array for the 40-pixel.

\begin{figure}[t]
\begin{center}
\includegraphics[width=1\linewidth, keepaspectratio]{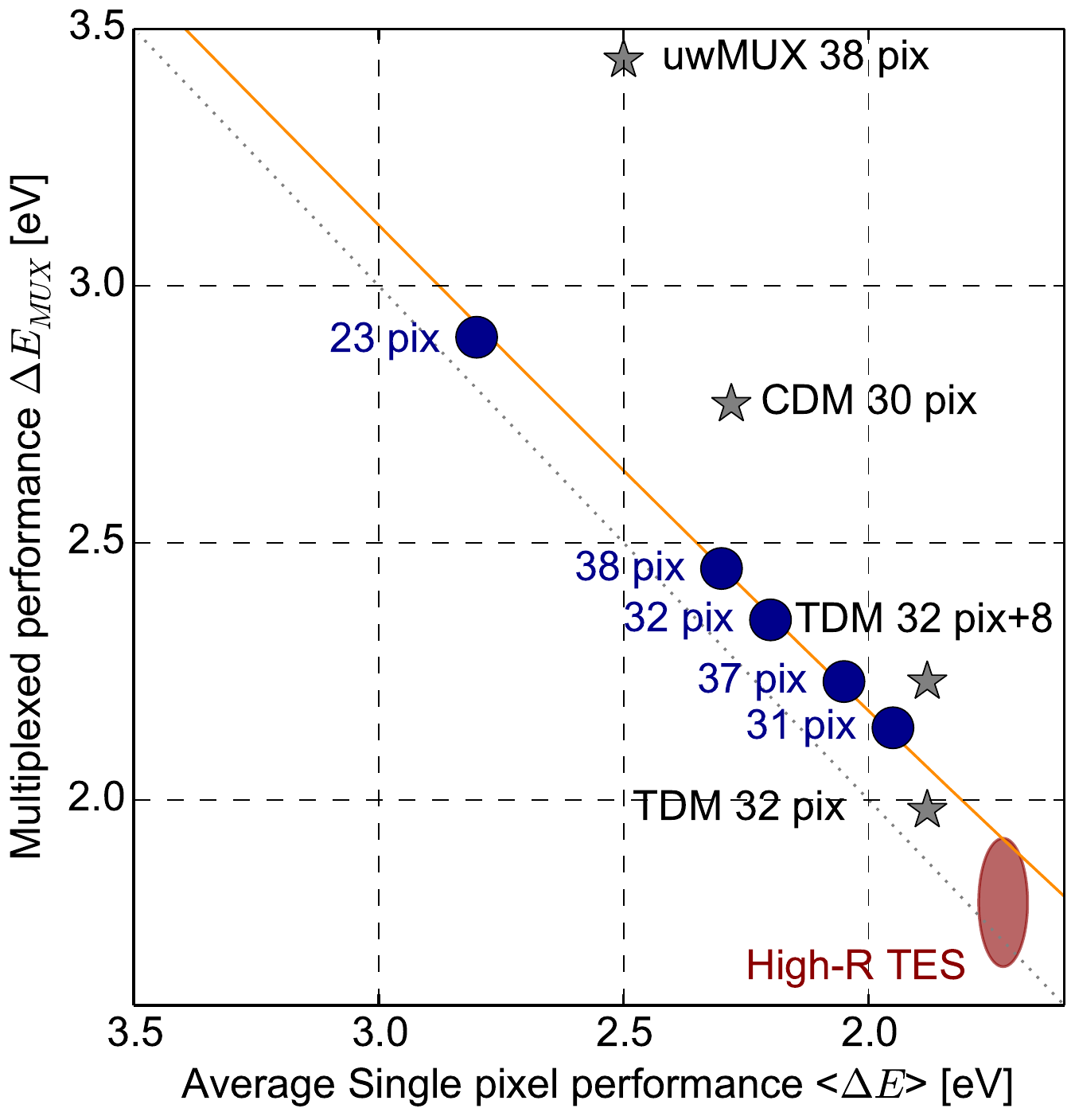}
\caption{\label{fig:comp}
Average of the single pixel performance ($\Delta E$)
and the multiplexed performance under FDM (blue circles) with results from other technologies (stars)
such as time-domain multiplexing (TDM) (32 TESs + 8 repeats of the last row, $\Delta E=2.23$ eV)\cite{durkin19}, code-division multiplexing (CDM) (30 pixel MUX, $\Delta E=2.77$ eV)\cite{morgan16},  and microwave SQUID multiplexing (umMUX) (38 pixel MUX, $\Delta E=3.44$ eV)\cite{nakashima20}.
The orange line indicates the estimated degradation $0.9$ eV in quadrature. The red oval indicates the expected range of the multiplexed performance with the high-$R$ device\cite{dewit20}.
}
\end{center}
\end{figure}

After the implementation of all mitigating measures discussed above, we performed a demonstration of the FDM  readout. The TESs were operated around 20\% of the transition. The  AC-bias was set to about 50\% of the maximum level, and the signal feedback to about 36\%. The crest factors were about 7 in both signals. The electrical cross-talk events were excluded by rejecting positive peaks in the TESs baseline. For both setups, one pixel was switched off because of exceptionally strong IMLN. 
The results with 31 pixels and 37 pixels are summarized in the table~\ref{tab:mux}.
The summed spectrum shows the spectral resolution at 5.9 keV of 
$\Delta E_{\rm 31 pix}$=2.14$\pm0.03$eV
and  $\Delta E_{\rm 37 pix}$=2.23$\pm$0.03 eV 
for the XFDM and the 40-pixel setups, respectively (Fig.\ref{fig:MUX_sum}).
Fig.\ref{fig:comp} shows the compilation of recent demonstrations of FDM readout and comparison with other multiplexing technologies. 
The degree of the degradation due to multiplexing is estimated to be $\Delta E_{deg}$=0.9 eV in the quadrature (as indicated by the orange line in Fig.\ref{fig:comp}).
We attribute the cause of the present degradation to be a combination of 1.) the residual of the AC Josephson  effect, 2.) sub-optimal thermalization in the array, and  3.) impact of the IMLN.
These issues, in particular on 2) and 3),  are not fundamental limitations and can be mitigated with incremental improvements such as the relocation of the AMP SQUIDs outside the detector temperature stage, the further optimising of the readout circuit., 
and the fabrication of TESs with even higher resistances.
Higher resistance devices in single pixel mode showed a better energy resolution of $\Delta E$=1.6-1.7 eV\cite{dewit20} (the red dotted line in Fig.\ref{fig:comp}).
Testing these devices and optimizations in multiplexed mode requires different cryogenic components that are not yet available (e.g., LC filters and transformers). 
Considering the well-described scaling relation shown in Fig \ref{fig:comp}, and the fact that the bottlenecks are identified, we see room for improvement in a future demonstration together with a better single-pixel performance device.

In summary: we have shown a demonstration of FDM multiplexed readout of 31 pixels from an 8$\times$8 TES array and 37 pixels from a 32$\times$32 TES array, both fabricated at SRON.
Our current best results are  $\Delta E_{\rm MUX 31 pixels}=2.14\pm0.03$ eV,
 $\Delta E_{\rm MUX 37 pixels}=2.23\pm0.03$ eV at 5.9 keV. The performance of these demonstrations is comparable to those reported by other multiplexing technologies (Fig.\ref{fig:comp}). 
This FDM readout technology will provide a means to increase the number of pixels in TES X-ray spectrometers for various fields from ground-based science to astronomy (X-ray\cite{hubs, sdios}, and from infrared to cosmic microwave background measurements\cite{hazumi20}).

\noindent 
\subsection*{Data availability}
The data that support the findings of this study are available from the corresponding or contributing authors upon reasonable request.

\begin{acknowledgments}
SRON Netherlands Institute for Space Research is supported financially by NWO, the Dutch Research Council. 
This work is funded partly by NWO under the research programme Athena with project number 184.034.002, and partly by the European Space Agency (ESA) under ESA CTP Contract No. 4000130346/20/NL/BW/os. 

\end{acknowledgments}

\bibliography{references}

\end{document}